\documentclass{article}
\usepackage[utf8]{inputenc}
\usepackage{authblk}
\usepackage{indentfirst}
\usepackage{hyperref}
\usepackage{longtable}
\usepackage{tabularray}
\usepackage{setspace}
\usepackage[margin=1in]{geometry}
\usepackage{graphicx}
\graphicspath{ {./figures/} }
\usepackage{subcaption}
\usepackage{amsmath}
\usepackage{lineno}
\usepackage{multirow}
\usepackage{color,soul}
\usepackage[version=3]{mhchem}

\usepackage[biblabel]{cite}

\title{Optimal pre-train/fine-tune strategies for accurate material property predictions}

\author[1]{Reshma Devi}
\author[2,*]{Keith T. Butler}
\author[1,*]{Gopalakrishnan Sai Gautam}

\affil[1]{Department of Materials Engineering, Indian Institute of Science, Bengaluru, 560012, India}
\affil[2]{Department of Chemistry, University College London, London WC1E 6BT, United Kingdom}
\affil[*]{Email: \href{mailto:k.t.butler@ucl.ac.uk}{k.t.butler@ucl.ac.uk}; \href{mailto:saigautamg@iisc.ac.in}{saigautamg@iisc.ac.in}}

\date{}

\begin{document}

\maketitle

\begin{abstract}
Overcoming the challenge of limited data availability within materials science is crucial for the broad-based applicability of machine learning within materials science. One pathway to overcome this limited data availability is to use the framework of transfer learning (TL), where a pre-trained (PT) machine learning model (on a larger dataset) can be fine-tuned (FT) on a target (typically smaller) dataset. Our study systematically explores the effectiveness of various PT/FT strategies to learn and predict material properties with limited data. Specifically, we leverage graph neural networks (GNNs) to PT/FT on seven diverse curated materials datasets, encompassing sizes ranging from 941 to 132,752 datapoints. We consider datasets that cover a spectrum of material properties, ranging from band gaps (electronic) to formation energies (thermodynamic) and shear moduli (mechanical). We study the influence of PT and FT dataset sizes, strategies that can be employed for FT, and other hyperparameters on pair-wise TL among the datasets considered. We find our pair-wise PT-FT models to consistently outperform models trained from scratch on the target datasets. Importantly, we develop a GNN framework that is simultaneously PT on multiple properties (MPT), enabling the construction of generalized GNN models. Our MPT models outperform pair-wise PT-FT models on several datasets considered, and more significantly, on a 2D material band gap dataset that is completely out-of-distribution from the PT datasets. Finally, we expect our PT/FT and MPT frameworks to be generalizable to other GNNs and materials properties, which can accelerate materials design and discovery for various applications.
\end{abstract}


\section{Introduction}
Machine learning (ML) based architectures play a pivotal role in materials research due to their high accuracy in predicting properties at low computational costs,\cite{xu2023small,chan2022application,du2023machine,xian2023machine} which can accelerate materials discovery for various applications. The accuracy of an ML model depends on the quantity and quality of data, the model framework, and the kind of algorithm used for training. Importantly, regression or classification models built on `simple' composition-based descriptors (that may be tailored with scientific intuition) typically underperform in material property predictions compared to models that take the full structural information as input, such as a graph neural networks (GNNs).\cite{witman2023defect} However, GNNs perform better than `simple' models only when the dataset size is large (i.e., $>$ 10$^{4}$ datapoints),\cite{dunn2020benchmarking} while typical materials-related datasets are small (a few thousand datapoints or fewer).

Usually, GNNs exhibit high variance or increased over-fitting when trained on small datasets, resulting in larger generalisation errors than simple models.\cite{xu2023small,omee2024structure,zhao2019pairnorm,dunn2020benchmarking} Although an obvious way to obtain better GNNs is to increase the dataset size, this may be challenging for specific properties that are difficult to compute or measure, such as defect formation energies, molecular adsorption energies on surfaces, ionic conductivities, electron-phonon coupling constants, and grain boundary energies, to name a few. Another pathway is to use models or frameworks that train well on small datasets, without necessarily exhibiting high variance. In the context of training robust models on small datasets, transfer learning (TL) as a strategy has recently gained immense popularity to improve model performance.\cite{george2018classification,kaur2020deep,liu2020deep,das2022automated} Specifically, TL allows knowledge transfer from a source domain, typically with a large dataset size, to a target domain of interest with a small dataset size.\cite{weiss2016survey} Usually, the parameters of selective (or all) layers of the model pre-trained (PT) on the source dataset are tuned or re-trained on the target dataset to make predictions on the target property, a process referred to as fine-tuning (FT).\cite{gupta2024structure,gupta2021cross,weiss2016survey}  Otherwise, parameters from the PT model can be used to construct feature vectors for a new deep learning (DL) model on a target property, a technique referred to as feature extraction.\cite{gupta2024structure} Note that the benchmark for a TL model is always to perform better than models trained from scratch (referred to as scratch models) on the smaller target dataset.

Several recent studies have sought to address the issue of small dataset size in materials science using TL. For example, Jha et al.\cite{jha2019enhancing} employed the ElemNet\cite{jha2018elemnet} architecture to TL and reduce the mean absolute error (MAE) in predicting experimental formation energies to 0.0731 eV (from 0.1325~eV in scratch models) by PT the same model on density functional theory (DFT\cite{kohn1965self,hohenberg1964inhomogeneous}) computed formation energies. The size of the PT and FT datasets were 341,000 and 1,643, respectively. Subsequently, Gupta et al.\cite{gupta2021cross} reduced the MAE on experimental formation energies further to 0.0708~eV by utilising cross-property TL and feature extraction. Notably, the DFT-calculated PT dataset in both the above works\cite{jha2019enhancing,gupta2021cross} came from the open quantum materials database (OQMD).\cite{saal2013materials,kirklin2015open}

Earlier Lee and Asahi\cite{lee2021transfer} included structural information in TL by using the crystal graph convolutional neural network (CGCNN)\cite{xie2018crystal} as the base architecture. By loading the weights of the PT model for FT on six different properties, including materials project (MP\cite{jain2013commentary}) formation energy and band gap datasets, the authors highlighted that the prediction accuracy of the FT model increased as the size of the PT dataset and/or the FT dataset increased. Subsequently, Gupta et al.\cite{gupta2024structure} used the atomistic line graph neural network (ALIGNN\cite{choudhary2021atomistic}) architecture, which also takes structural information as input, to FT a model PT on MP formation energy on multiple properties from the joint automated repository for various integrated simulations (JARVIS\cite{choudhary2020joint}) database, including JARVIS-3D, JARVIS-2D, etc. The authors\cite{gupta2024structure} reported that the feature extraction strategy generated better TL models on 54\% of instances compared to scratch models.

Chang et al. proposed a framework called the mixture of experts (MOEs\cite{chang2022towards}) to overcome limitations of TL, such as negative transfer\cite{wang2019characterizing} and catastrophic forgetting.\cite{chen2019catastrophic,kirkpatrick2017overcoming} The former refers to the case where the TL model performs worse than the scratch model and the latter refers to the case where the TL model overfits the target property due to loss of information captured from the PT model. The MOE model extracted features from a PT CGCNN model, using trainable gated functions, and was benchmarked on 19 material property regression tasks. Remarkably, the MOE model showed better performance on all 19 tasks compared to pair-wise TL models that used computational formation energies for PT. Additionally, Chang et al. also demonstrated that the extent of improvement in performance of the TL models (versus scratch models) varied non-monotonically with target dataset sizes. Importantly, there has been no study, until now, on how the choice of PT and FT dataset(s) and associated hyperparameters affect the generalisation ability of a new FT model. Another aspect that has not been rigorously explored in literature so far is the performance of a FT model that has been PT simultaneously on several different materials properties.

In this work, we systematically explore the efficacy of pair-wise and multi-property PT (MPT) approaches for TL in materials science datasets using the ALIGNN architecture as the base. We choose seven different properties from the Matminer library,\cite{ward2018matminer} including DFT average shear modulus (GV), frequency of the highest optical phonon mode peak (PH), DFT band gap (BG), DFT formation energy (FE), computed piezoelectric modulus (PZ), computed dielectric constant (DC), and experimental band gap (EBG). First, we optimise hyperparameters for pair-wise TL, such as PT and FT dataset size, and possible FT strategies, and examine the TL performance trends among PT and FT properties that are (not) related. Subsequently, we utilitise a MPT approach, where we PT on multiple properties simultaneously followed by FT on a target property, and compare its performance to scratch and pair-wise models. Our MPT approach is different from the MOE or feature extraction strategies, since we use the entire PT model in FT.  Apart from demonstrating that our MPT strategy outperforms the pair-wise TL models on 5/7 instances (in terms of MAE), we also show our MPT model to FT quite well on a completely out-of-distribution dataset, namely, the JARVIS-DFT 2D materials band gaps.\cite{choudhary2020joint} Also, we find our TL models to exhibit lower (or similar) MAEs, often utilising two or three orders of magnitude lower dataset sizes, than previous TL models.\cite{lee2021transfer,jha2019enhancing,gupta2024structure,chang2022towards} Finally, our work reveals robust PT/FT strategies for efficient TL between material property domains, which should further accelerate property predictions and materials discovery for various applications.

\section{Methods}
\subsection{Graph neural network}
\begin{figure}[t!]
\centering
\includegraphics[width=\linewidth]{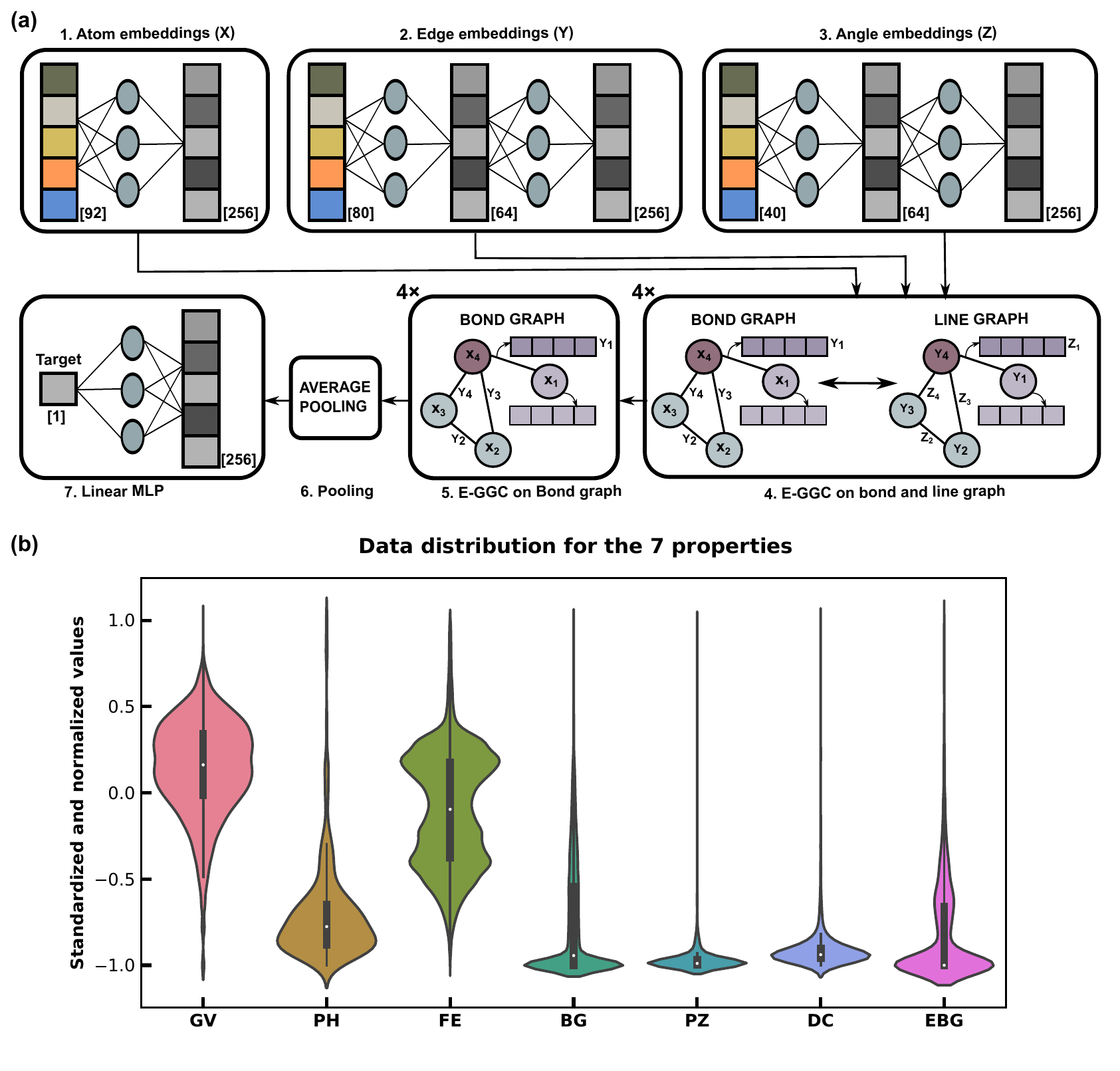}
\caption{a) Schematic describing the ALIGNN architecture. Each block corresponds to one layer of the model. The dimension of the embeddings is given in the bottom right corner within each layer. b) The standardized and normalized distributions of the seven datasets represented in the form of violin plots. The length of the black box inside each violin summarizes the interquartile range of the
corresponding data, while the white circle within each box indicates the median.}
\label{fig:alignn-data}
\end{figure}

GNNs offer a natural way to model molecules and solids: the nodes and edges in the graph correspond to the atoms (or molecules) and the interactions (or bonds) between them, respectively. Thus, GNNs capture the inherent connectivity among atoms/molecules and their local environment, which typically leads to better property predictions. Different GNN architectures have been proposed in the literature, such as CGCNN\cite{xie2018crystal} and its improved version (iCGCNN\cite{park2020developing}), materials graph neural network (MEGNet),\cite{chen2019graph} crystal Hamiltonian graph neural network (CHGNet),\cite{deng2023chgnet} and SchNET.\cite{schutt2017schnet} We use ALIGNN (v2023.04.01) in this work as it has been shown to achieve high-performance on materials property predictions and to generalise quite well out-of-distribution.\cite{omee2024structure}

Figure~{\ref{fig:alignn-data}a} depicts the ALIGNN architecture consisting of seven layers in total, beginning with initial layers (1, 2, and 3) that convert structural information into atom (X), bond (Y), and bond angle (Z) embeddings. X, Y, and Z embeddings serve as inputs to the N (layer 4), and M (layer 5) layers of edge-gated graph convolutions (E-GGC,\cite{dwivedi2023benchmarking}). Layers 4 and 5 are usually referred to as ALIGNN layers. Subsequently, global average pooling (layer 6) aggregates node information, which finally passes through a single fully connected prediction layer (layer 7). Note that ALIGNN includes bond angle information by using two crystal graphs, namely, an atomistic bond graph (or two body layers), and a line graph (three body layers). Nodes and edges in the atomistic bond graph represent atoms and bonds, while in the line graph they correspond to bonds and bond angles, respectively. The line graph is derived from the bond graph, and the updates to the edges and nodes in both graphs are obtained via E-GGC. Detailed information on the ALIGNN architecture and the default model configuration that we used can be found in prior work \cite{choudhary2021atomistic} and Table~S1 of the Supporting Information (SI). 

\subsection{Dataset description}
The datasets we have chosen in this study, which combine both computational and experimental quantities, are described below. Figures~S1-S7 of the SI illustrate the distribution of the crystal systems in each dataset. Additionally, Table~S2 compiles the maximum, minimum, standard deviation, and average values of each dataset.

\begin{enumerate}
    \item GV: the average shear modulus for 10,987 materials, computed using DFT and sourced from the MP database.
    \item PH: the highest frequency of the optical phonon mode peak (in units of cm$^{-1}$) for 1,265 materials, obtained from DFT calculations of Petretto et al.\cite{petretto2018high}
    \item FE: the DFT formation energy for 132,752 materials collected from the MP database.
    \item BG: the DFT-calculated band gap for 106,113 structures sourced from the MP database. The band gaps are calculated at the Perdew-Burke-Ernzerhof level of electronic exchange-correlation.\cite{perdew1996generalized}
    \item PZ: the piezoelectric modulus for 941 structures, computed through DFT calculations by Jong et al.\cite{de2015database} PZ represents the smallest dataset among those considered in this work.
    \item DC: the average eigenvalues of the total contributions to the dielectric tensor for 1,056 structures, as calculated by Petousis et al.\cite{petousis2017high}
    \item EBG: the experimental band gap data for 4,604 structures, compiled by Kingsbury et al.\cite{kingsbury2022performance}
\end{enumerate}

\subsection{Dataset cleanup} 
\begin{figure}[t!]
\centering
\includegraphics[width=\textwidth]{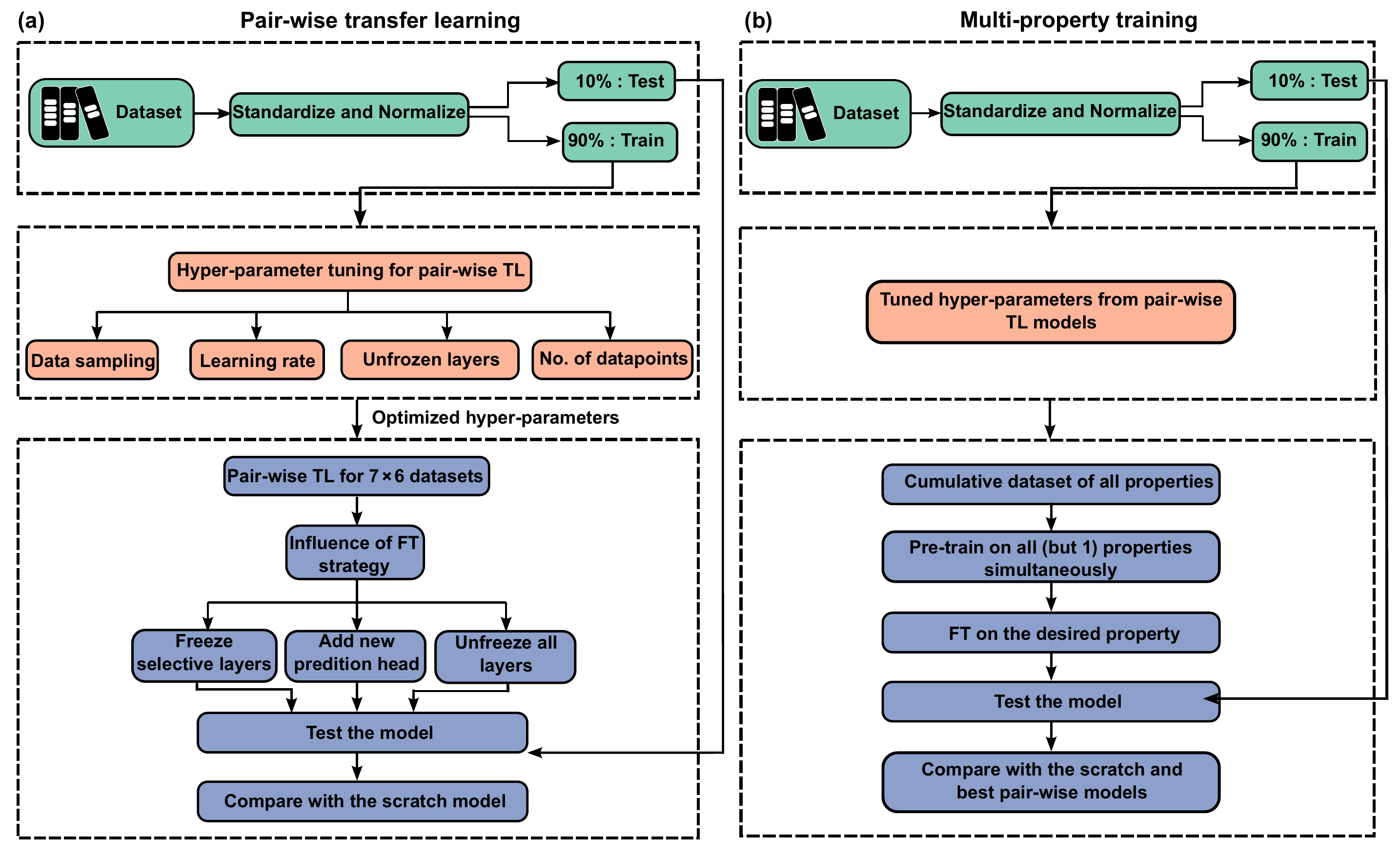}
\caption{a) Schematic illustrating the workflow (panel a) for pair-wise TL for the seven properties of interest, and (panel b) for MPT.}
\label{fig:flowchart}
\end{figure}

In order to ensure uniformity when comparing data reported in different scales or units, we standardized and normalized all values within each dataset considered. Figure~{\ref{fig:alignn-data}b} displays the distribution of the standardized and normalized values within each dataset as violins. Note that the BG, PZ, DC, and EBG follow a log-normal distribution compared to the GV, PH, and FE datasets. We used the standardized and normalized values for all PT and FT experiments through this work.

Figure~{\ref{fig:flowchart}a} describes the workflow of pair-wise TL among the seven datasets. Each dataset is split randomly into training and testing samples in a ratio of 90:10. The test dataset is never used in any of the PT or FT stages in pair-wise TL, either for training or validation. We further split the training data in the ratio 90:10 for training and validation for all the PT-FT experiments. We report and use the R$^{2}$ score and MAE on the test dataset to gauge the model performance in the following sections. The distribution of datapoints among the train and test splits for each dataset is compiled in Figures~S8-S11 of the SI.

First, we construct ALIGNN models that are trained individually on each of the seven datasets (of different sizes from the 90\% training data), which signify our PT models. Subsequently, we test the models on the corresponding 10\% test dataset, which represent the R$^2$ scores and MAEs of our scratch models. Finally, we FT each PT model on the remaining six datasets, leading to 7$\times$6 pairs of PT-FT models, which are denoted by `PT-FT'. For example, BG-FE implies that the model was PT and FT on BG and FE datasets, respectively. Where relevant, we specify the dataset size used for PT in a pair-wise model as `PT(size)-FT'. Thus, BG(1K)-DC refers to a model PT on the BG dataset with 1,000 datapoints and subsequently FT on the DC dataset.  

\subsection{Fine-tuning strategies}
\begin{figure}[t!]
\centering
\includegraphics[width=\linewidth]{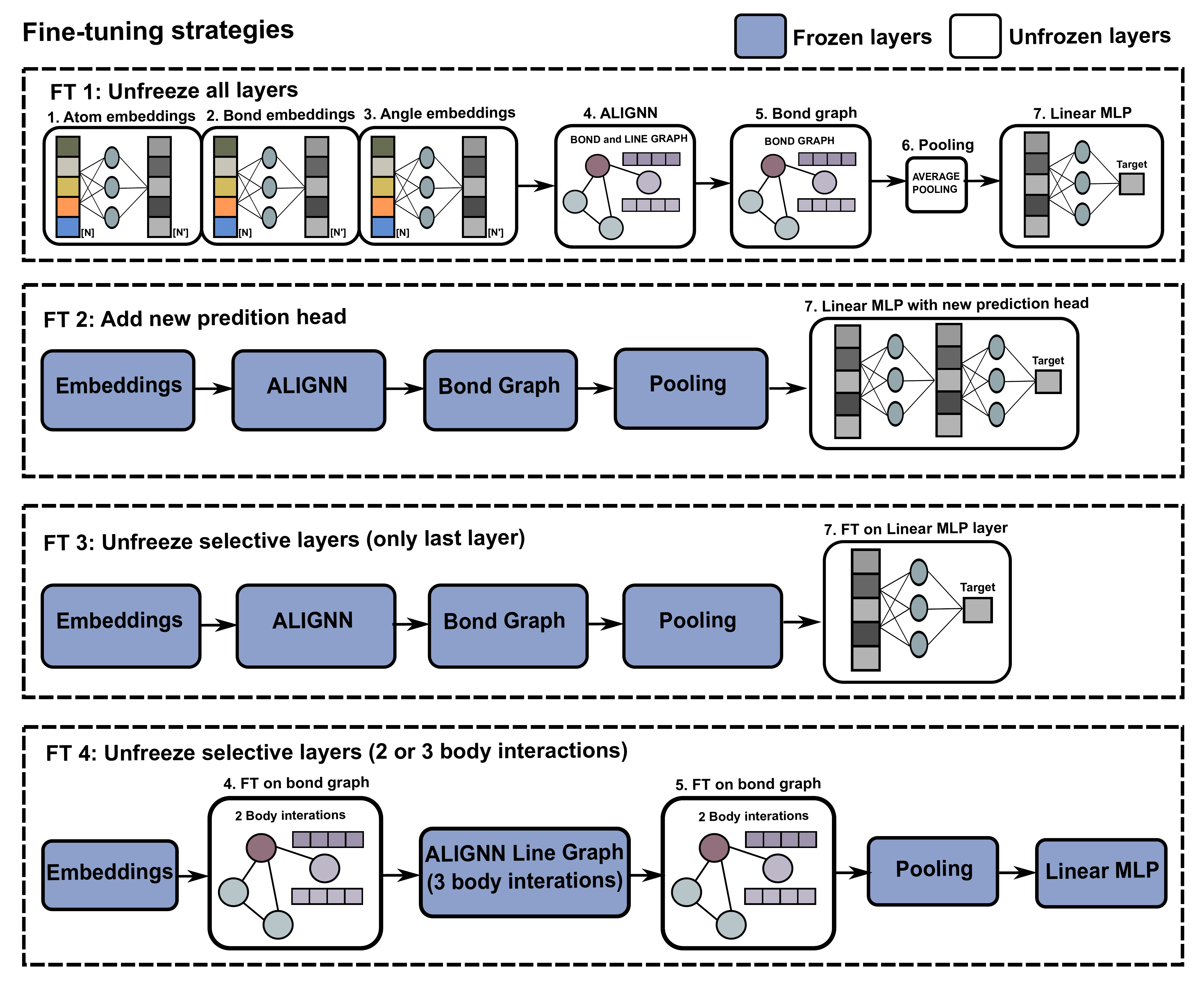}
\caption{The four FT strategies explored in this work. Blue boxes indicate frozen layers, and unfrozen layers are indicated by black outlined boxes. The contents within each unfrozen layer is indicated in each box, where the notations used within unfrozen layers are identical to Figure~{\ref{fig:alignn-data}}a.}
\label{fig:strategy}
\end{figure}

We have tested four different strategies of FT that are possible in a GNN-based architecture, as illustrated in Figure~{\ref{fig:strategy}} and described below. The parameters of the PT model that are kept fixed (re-computed) during FT are referred to as frozen (unfrozen) layers, as represented by solid blue (black outlined) boxes in Figure~{\ref{fig:strategy}}. The layer-number nomenclature for each FT strategy in Figure~{\ref{fig:strategy}} is identical to Figure~{\ref{fig:alignn-data}a}. 

\begin{itemize}
    \item FT strategy 1: Unfreeze all layers of ALIGNN
        \newline
        All the parameters in the PT model's seven layers are loaded as initializations and subsequently allowed to be re-computed during FT. Thus, this strategy gives the maximum degree of freedom available for the model to update itself during FT.
    \item FT strategy 2: Add a new prediction head  
        \newline
        A new multi-layer perceptron (MLP) layer is added before the prediction head. The parameters of the other six layers of the PT model are fixed apart from the final modified layer. Hence, only the fully connected linear MLP introduced before the prediction head is allowed to re-train on the target property.
    \item FT strategy 3: Unfreeze only the last layer 
        \newline
        This is the conventional idea of re-training only the last layer of a DL model in TL. Thus, we re-train only the final layer of the ALIGNN architecture keeping the parameters of the other six layers fixed. This strategy provides the least degree of freedom for the model to update itself during FT. 
    \item FT strategy 4: Unfreeze selective (interaction) layers 
        \newline
       The two body interaction layers (bond graphs) corresponding to the 4$^{th}$ and the 5$^{th}$ layers of ALIGNN or the three body interaction layers (line graphs) within the 4$^{th}$ layer are allowed to be unfrozen while keeping the rest of the model constant.
\end{itemize}

\subsection{Hyperparameter tuning}
\label{subsec:hyperparameter}
Apart from the FT strategy, there are other important hyperparameters (listed below and illustrated in second block of Figure~{\ref{fig:flowchart}a) which need to be optimized for both pair-wise and MPT TL. Given the large number of PT-FT combinations that can be created among the pair-wise models for optimizing hyperparameters, we chose the following set of PT-FT pairs: BG-FE, FE-BG, DC-BG, and BG-DC. The choice of the above PT-FT pairs was motivated by $i$) the presence or absence of physical correlation between properties (e.g., DC-BG are correlated but FE-BG are not), $ii$) the difference in data distribution between PT and FT sets (e.g., FE is bimodal while BG is log-normal), and $iii$) the inclusion of the largest two datasets (BG and FE). 

We used 90\% of the full dataset, which is further split 90:10 for training and validation, for PT in all cases. We fixed the FT dataset size to 500, and chose the conventional FT strategy 3 (Figure~{\ref{fig:strategy}}) to optimize the hyperparameters, unless otherwise specified. Note that we used the set of optimized hyperparameters from this exercise for MPT TL as well. The details of each hyperparameter optimization is compiled in Section~S3 of the SI (see Tables~S3-S6). After optimizing hyperparameters, we performed five different (random) sampling trials for each TL experiment, with average values used for all illustrations and margins of errors for confidence intervals of 95\% reported in the SI.

\textbf{Data sampling}: The way the available data is sampled during FT can play a role in the model performance, since we capped the FT datasize to 500, which is fewer than the smallest dataset that we have considered (i.e., PZ with 941 datapoints). We selected the 500 FT datapoints by random, weighted, and uniform sampling for the PT-FT pairs mentioned above. Importantly, we identify random sampling to perform better than the other two techniques in all PT-FT pairs except BG-FE (see Figure~S12a).

\textbf{Learning rate}: To estimate the optimal learning rate for FT, we evaluated the model performance (quantified by R$^2$ scores) for select PT-FT pairs with four different learning rates (10$^{-2}$,10$^{-3}$,10$^{-4}$, and 10$^{-5}$). We used random sampling and model configuration as tabulated in Table~S4. Notably, we observe that a higher learning rate (10$^{-2}$ or 10$^{-3}$) offers better model performance (see Table~S5), which may be attributed to the need for greater re-training of parameters in our tasks than in vision and language tasks.\cite{kim2022transfer,chronopoulou2019emb} 

Among the higher learning rates, we find a rate of 10$^{-3}$ to be marginally better than 10$^{-2}$ owing to better convergence and lower noise in validation R$^{2}$ scores (see Figure~S13). To verify that higher learning rates provide better performance even on changing the FT strategy, we employed strategy 2 (from Figure~{\ref{fig:strategy}}) for three different learning rates (10$^{-3}$,10$^{-4}$, and 10$^{-5}$). Importantly, we find similar trends of high R$^{2}$scores at high learning rates (see Figure~S12b). Hence, we fix the optimal learning rate as 10$^{-3}$ for all subsequent experiments.

\textbf{Number of frozen layers}: Changing the number of frozen layers during TL should impact how well the model re-trains on the FT dataset. To explore this, we varied the number of frozen layers within the ALIGNN architecture to be either 1 (the embedding layer) or 6 (until the final layer), which represent two extreme scenarios. Additionally, we added a new prediction head in both cases (similar to FT strategy 2). Importantly, we find that the model performs better on the FT dataset with higher number of unfrozen layers (see Figure~S12b), suggesting that the PT model requires significant updating to accurately perform the FT task. 

\textbf{Number of datapoints}: To examine the influence of the size of the FT dataset, we varied the FT dataset size from 500 to 1000. Expectedly, we observe that the R$^{2}$ scores increase as the FT dataset size increases (Table~S6). We have included a more detailed discussion on the influence FT dataset size in Section~3.1.  

\subsection{Multi-property training}
Given that pair-wise TL using specific PT-FT pairs can be quite specific to the property that they are trained for, we construct a more general PT model involving multiple properties, similar to the multi-task learning model proposed by Sanyal et al.\cite{sanyal2018mt} While MPT has demonstrated better FT in material property predictions,\cite{sanyal2018mt} the number of properties used in PT was small (2-3 properties). Here we build an extensive MPT dataset considering seven different prediction targets, by agglomerating all the seven individual datasets considered in this work, to yield a cumulative dataset of 132,270 points. The workflow used in constructing the MPT model is given in Figure~{\ref{fig:flowchart}}b and the model configuration is specified in Table~S7.

For each data point (i.e., structure), we associate a one-hot encoded vector and a property list vector, each with a dimension of seven. The former describes if a particular property value is available for a structure, and the latter gives the respective value of that property. We define a multi-property loss function (per structure), as in Eq.~{\ref{eq:loss}}, where $N$ is the number of properties, $y_p$ and $y_t$ are the predicted and target property values, $i$ is the property index, and $\delta$ is the one-hot vector entry per property. 

\begin{equation}
\label{eq:loss}
\mathcal{L} = \frac{1}{N}\sum_{i=1}^{N} |y_p^i - y_t^i|\delta^i
\end{equation}

To FT a model on a specific property (e.g., BG), we train a single MPT model simultaneously on the remaining six properties (i.e., FE, GV, PH, PZ, DC, and EBG). Note that we filter out the property information from all datapoints that are used in the FT process from PT so that the MPT model is not exposed to any of the FT datapoints. For instance, to FT on BG, we modify the one-hot vector entry of all datapoints in the cumulative dataset that contain a BG to zero, so that the MPT model does not PT on any BG information. During PT, the embedding obtained from the graph convolutions of ALIGNN (i.e., after layer 6) is passed to fully connected individual MLPs dedicated to each of the six PT properties considered. After PT, the MTP model is FT by adding two extra layers of MLP to the PT model, before the prediction head, and re-training the entire configuration on the desired target property.

\section{Results}
\subsection{Influence of FT dataset size}
\begin{figure}[h]
\centering
\includegraphics[width=\linewidth]{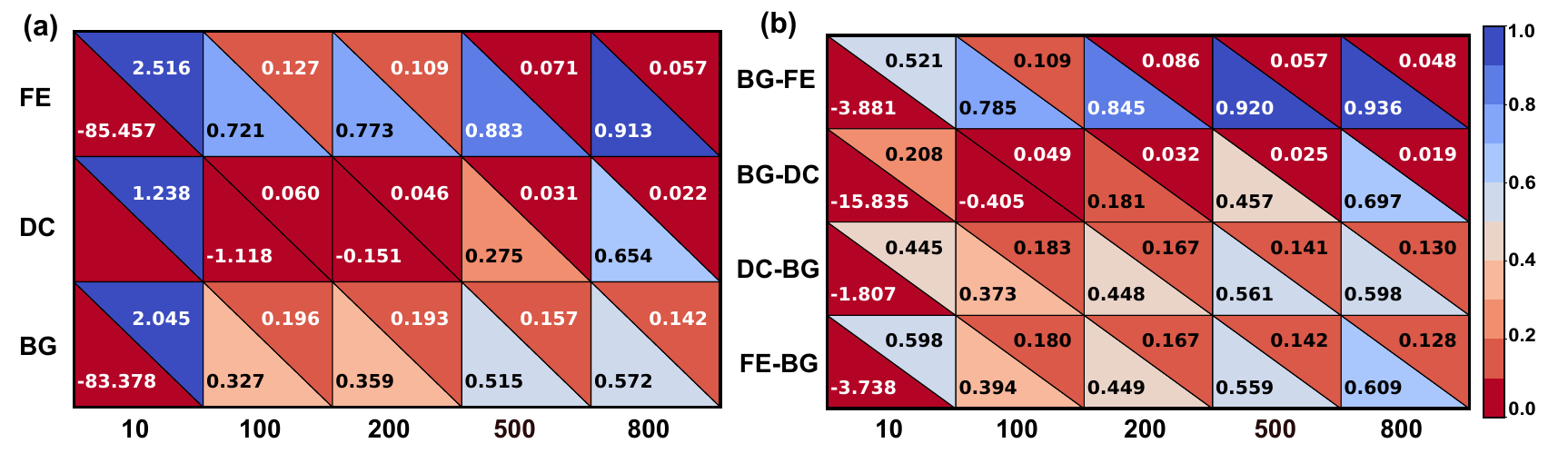}
\caption{Test R$^{2}$scores (lower triangles) and MAEs (upper triangles) for the models constructed from (panel a) scratch and (panel b) for select PT-FT pairs. The $x$ and $y$ labels in both panels correspond to the dataset size and name or PT-FT pair, respectively. }
\label{fig:heatmap-1}
\end{figure}

Figure~{\ref{fig:heatmap-1}a} illustrates the heatmaps for the scratch models for the FE, DC, and BG datasets for varying dataset sizes, namely, 10, 100, 200, 500, and 800. The $y$ labels in Figure~{\ref{fig:heatmap-1}a} denote the dataset name, while the $x$ labels denote the corresponding training dataset size. The lower and upper triangles represent R$^2$ scores and MAEs within the test dataset, respectively. The color code of the heatmap varies between red and blue, corresponding to the values 0 and 1. Thus, a good model has high R$^{2}$scores (close to 1 or blue triangles) and low MAEs (close to 0 or red triangles). The test R$^2$ scores and MAEs of scratch models for all datasets (and corresponding dataset sizes from 10 to 800) are compiled in Tables~S8 and S9 of the SI.

Figure~{\ref{fig:heatmap-1}b} displays the heatmaps for pair-wise TL models, where we performed PT-FT on select pairs, namely, BG-FE, BG-DC, DC-BG, and FE-BG. The notations used in Figure~{\ref{fig:heatmap-1}b} are similar to panel a, with the $y$ labels in panel b indicating the specific PT-FT pair. Note that we capped the FT dataset size to 800 as it is roughly 90\% of the smallest dataset size that we have considered (i.e., PZ with 941 datapoints). Similarly, for PT among all PT-FT pairs, we capped the dataset size to 941, so as to remove the influence of PT dataset size. We used model parameters as listed in Table~S1 and used strategy 1 for FT (Figure~{\ref{fig:strategy}}). All PT-FT experiments were conducted for five different random trials, and the mean results are plotted in Figure~{\ref{fig:heatmap-1}}b. 

The R$^{2}$scores and MAEs of the FT models are better than the scratch models for all four PT-FT pairs considered (Figure~{\ref{fig:heatmap-1}}). For example, the R$^{2}$score and MAE for BG800 (i.e., scratch BG model with an 800 point dataset for training) are 0.572 and 0.142, respectively (Figure~{\ref{fig:heatmap-1}a}). In comparison, the FE-BG800 model (i.e., FE as PT dataset and an 800 point dataset for FT on BG) exhibits R$^2$ and MAE of 0.609 and 0.128, respectively. Similarly, we observe the DC-BG800 model to perform better than the scratch model as well (R$^2$ and MAE of 0.598 and 0.130, respectively). Overall, we observe improvements in both R$^{2}$scores and MAEs in all three datasets with PT-FT models compared to scratch. Also, the PT-FT models exhibit better (or similar) performance compared to scratch at smaller dataset sizes (except for size 10).

The R$^2$ and MAE improve for a PT-FT model with an increase in FT dataset size, which is expected. For example, BG-FE800 has a better R$^2$ (0.936) and MAE (0.048) than BG-FE500 (0.920 and 0.057, respectively). Note that the percentage improvement in R$^{2}$scores and MAEs saturate as the FT dataset size increases. For instance, the percentage improvement in R$^{2}$score for BG-FE on increasing the dataset size from 200 (R$^2$=0.845) to 500 (R$^2$=0.920) is $\sim$8.9\%, whereas it is only $\sim$1.7\% when it is increased to 800 (R$^2$=0.936) from 500. Thus, the choice of the FT dataset size should be preferably close to the point where the R$^2$ scores and MAEs saturate (i.e., around 800 datapoints in Figure~{\ref{fig:heatmap-1}}). We, therefore, fixed the FT dataset size to 800 for all the following experiments. The results for the FT dataset sizes of 10, 100, 200, and 500 for the following sections are illustrated in Tables~S10-S20 of the SI. 

\subsection{Influence of PT dataset size}
\label{sec:PT-size}
We chose the two largest datasets in our consideration, FE and BG, to study the influence of PT dataset size, and utilized 1K, 5K, 10K, 50K, and 100K randomly sampled subsets from the 90\% train datasets of FE and BG. Using FT strategy 1, we performed FT on a fixed 800-point dataset, while varying the PT dataset sizes. Panels a and b of Figure~{\ref{fig:ld-radar}} show the percentage change in the test R$^2$ scores of the PT-FT models with respect to the scratch models, where we define the percentage change as [(R$^2$ of PT-FT)-(R$^2$ of scratch)]$\times$100/(absolute value of R$^2$ of scratch). Thus, positive (negative) values of percentage changes indicate better (worse) performance of the PT-FT models versus scratch. 

\begin{figure}[h]
\centering
\includegraphics[width=\linewidth]{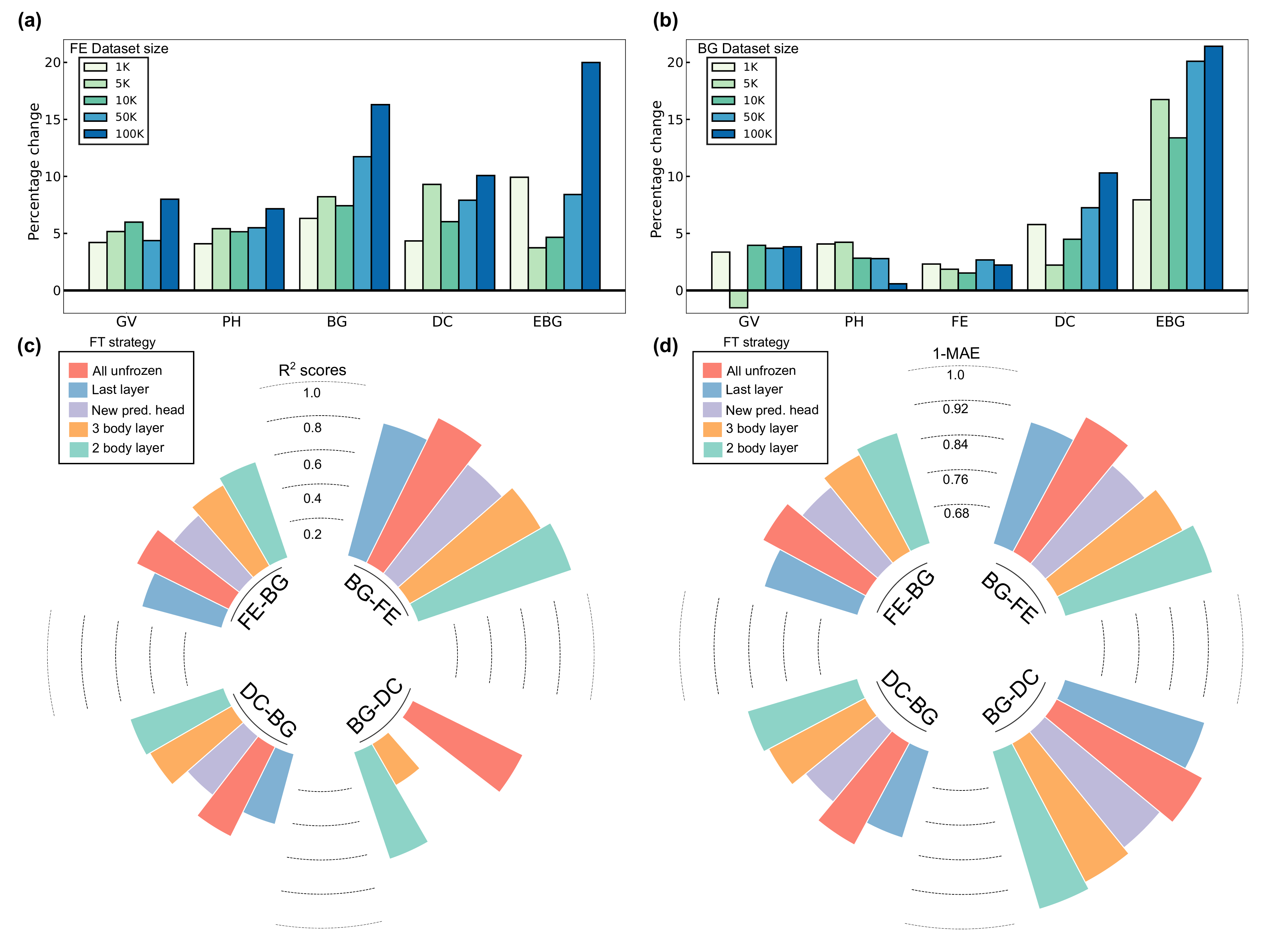}
\caption{Percentage change in test R$^2$ scores of PT-FT models with respect to corresponding scratch models, with the PT done on FE (panel a) and BG (panel b). The FT datasets are indicated along the $x$ axis, while the FT dataset size is kept to 800. PT dataset sizes are indicated with different colored bars. Circular bar plots illustrating the performance in terms of R$^{2}$scores (panel c) and 1-MAEs (panel d) for select PT-FT pairs. Different FT strategies are illustrated by the different colored bars. The range of the bars are indicated by the text annotations provided across the concentric circles in both panels.}
\label{fig:ld-radar}
\end{figure}

The performance of models PT with FE and BG are displayed in panels a and b of Figure~{\ref{fig:ld-radar}}, respectively, with the varying PT dataset sizes indicated by different bar colors in both panels. Data from the FE-PZ and BG-PZ pairs are plotted in Figure~S15a for ease of visualization, since the percentage improvements in PT-FT models in these pairs are one order of magnitude higher than the other PT-FT pairs considered. Tables~S10-S13 of the SI tabulate the R$^2$ scores and MAEs for the FT dataset sizes of 10, 100, 200, and 500, while the PT dataset size (i.e., for FE and BG) is also varied, while Tables~S14 and S15 compile the percentage changes in R$^2$ scores and MAEs. Figures~S16 and S17 visualize the data compiled in Tables~S14 and S15.

The percentage change in performance of PT-FT models versus scratch models is non-monotonic for both the PT datasets (FE and BG), as the dataset size increases. For example, in FE-BG, FE-DC, FE-PH, BG-FE, and BG-EBG pairs, the PT(10K)-FT models show lower improvement versus scratch compared to PT(5K)-FT and PT(50K)-FT models. In case of PT with FE, the FE(100K)-FT offers the best improvement in R$^2$ scores for all FT datasets (Figure~{\ref{fig:ld-radar}}a and Figure~S15a), with the FE(50K) model being the next best in all FT cases except GV. The improvement in performance with more PT data can be attributed to the GNN learning a better representation of the `normal' data distribution of FE (Figure~{\ref{fig:alignn-data}}b), which facilitates FT on a newer property. At smaller PT dataset sizes (say $\leq$ 10K points), there is a possibility that the GNN gets strongly optimized to a smaller class of structures/chemistries and lacks generalization. Thus, with a normal distribution, increasing the amount of PT data available ($\geq$ 50K points) helps in obtaining better models, while the models will exhibit some non-monotonicity in performance at small dataset sizes ($\leq$ 10K).  

In contrast to FE, including larger amount of data for PT with BG does not always result in better FT model performance. For example, BG-GV, BG-PH, BG-PZ, and BG-FE pairs exhibit poorer improvement versus scratch when exposed to 100K PT (BG) datapoints compared to 50K (Figure~{\ref{fig:ld-radar}}b and Figure~S15a). In contrast, for DC and EBG, PT with BG(100K) gives the best performance upon FT. Note that BG, DC, and EBG are correlated properties as well as follow a log-normal distribution, while GV, PH, and FE exhibit a normal distribution (Figure~{\ref{fig:alignn-data}}b). Thus, combining the trends observed on PT with FE and BG, we can conclude that a larger amount of PT data is only helpful if the PT data distribution is normal (e.g., FE),  or if the FT is done on a target that is correlated with the PT data (e.g., BG-EBG). Additionally, we observe that the similarity in data distribution in both the PT and FT datasets is a weak handle in determining whether performance improves with including more PT data, as shown by BG(50K)-PZ displaying better performance than BG(100K)-PZ, where both BG and PZ follow a log-normal distribution. For the following sections, we identify 100K and 50K to be the best dataset sizes to use while employing FE and BG datasets for PT, respectively.

\subsection{Best FT strategy}
To determine the best FT strategy, we performed a comparison among the four strategies considered (Figure~{\ref{fig:strategy}}) for controlled PT and FT dataset sizes of 941 and 800, respectively, and select PT-FT pairs (BG-FE, BG-DC, DC-BG, and FE-BG), as illustrated in Figure~{\ref{fig:ld-radar}}b. We plot R$^2$ scores and 1-MAEs in panels c and d of Figure~{\ref{fig:ld-radar}}, respectively, in the form of circular bars. The ranges of the bars are indicated by numerical notations across the concentric circles. The circular bars that are farther from the origin represent both better R$^2$ scores and MAEs. The unfreezing of two body and three body interaction layers corresponding to FT strategy 4 are individually represented.

Importantly, we observe that unfreezing all the layers (or FT strategy 1) offers the best performance in all the PT-FT cases, with respect to both R$^{2}$scores and MAEs in contrast to generally applied PT/FT strategies, where part of the model is frozen for FT. This indicates that the FT requires significant amount of re-training for the models to become generalized enough on the FT property. The performance of FT strategy 1 is followed by FT strategy 4 (unfreezing two body layers followed by three body layers). The good performance of FT strategy 4 is a further indication that re-training of several layers is required for accurate FT and suggests that the majority of the model performance in ALIGNN is being governed by the bond graph layers followed by the line graph layers. Overall, we identify FT strategy 1 to be the best performer among the strategies considered in this work. 

\subsection{Pair-wise TL for 7$\times$6 combinations}
\begin{figure}[p]
\centering
\includegraphics[width=\linewidth]{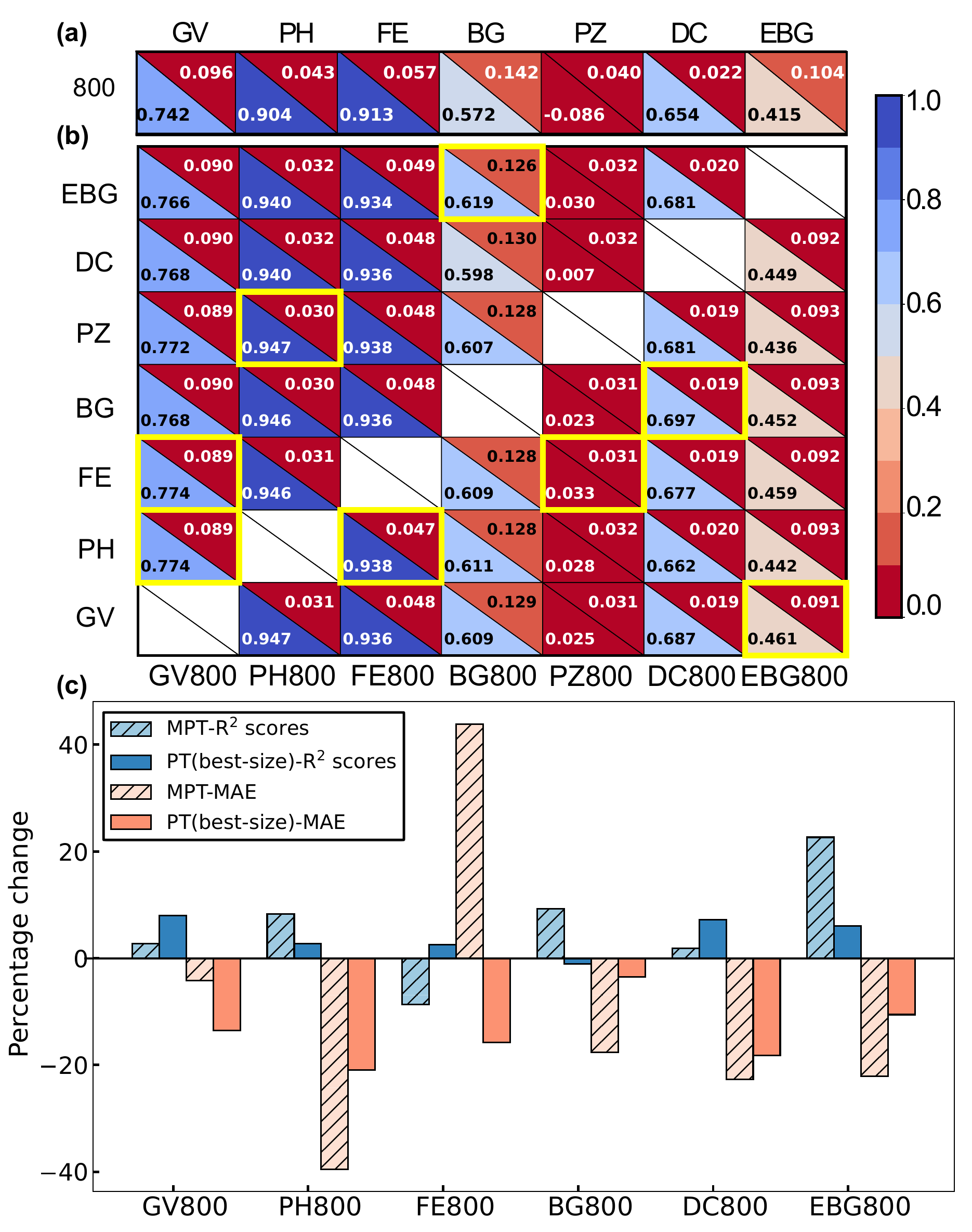}
\caption{a) R$^{2}$scores (lower triangles) and MAEs (upper triangles) for the models corresponding to the seven datasets constructed from scratch. The $x$ and $y$ labels represent the dataset name and size. b) Performance of pair-wise TL for the 7$\times$6 combinations. The $y$ labels represent the PT datasets, and the $x$ labels display the FT datasets and sizes (i.e., 800 datapoints for all FT). The best performing models are highlighted by yellow boxes. c) Percentage change in R$^2$ scores (blue bars) and MAEs (orange bars) of the MPT (hashed bars) and PT(best-size)-FT (solid bars) models with respect to scratch.}
\label{fig:heatmap-2}
\end{figure}

By keeping the PT and FT sizes as 941 and 800 (corresponding to our smallest dataset size) and the FT strategy as 1, we analyze the performance of pair-wise TL on all 7$\times$6 PT-FT combinations that are possible among the seven datasets considered in this work. Figure~{\ref{fig:heatmap-2}a} and {\ref{fig:heatmap-2}b} display heatmaps containing the test R$^{2}$scores (lower triangles) and test MAEs (upper triangles) for all the seven scratch models (for dataset size of 800) and the 7$\times$6 combinations, respectively. The performance of the FT models (panel b in Figure~{\ref{fig:heatmap-2}}) are compared against scratch models (panel a), i.e., the comparison is against metrics along each column of the two panels (see Table~S18 for percentage change in R$^2$ and MAE of PT-FT vs. scratch models). The yellow highlighted boxes indicate the best pair-wise models (highest R$^2$ and lowest MAE) for each FT dataset. The margin of error for each PT-FT combination after the five random trials are tabulated in Tables~S16 and S17, which also includes data for different FT dataset sizes apart from 800 (namely, 10, 100, 200, and 500). Figure~S18 visualizes the data compiled in Tables~S16 and S17.  

We can observe from  Figure {\ref{fig:heatmap-2}b} that the TL models outperform the scratch models in all FT cases (with 800 datapoints), with an average percentage increase in R$^{2}$ score and MAE of 25.15\% and 16.42\%, respectively. Additionally, the pair-wise models achieve an equivalent performance at fewer datapoints compared to scratch models at trained on larger dataset sizes (see Figure~S19 for a compilation). For example, the R$^{2}$score and MAE for the scratch model GV800 are 0.742 and 0.096, respectively (Figure~{\ref{fig:heatmap-2}}a). We obtain an equivalent performance with the PH-GV500 model, which exhibits a similar R$^{2}$score and MAE of 0.722 and 0.099 while using 300 fewer datapoints than scratch (Figure~S18e and S19). 

The best PT model for each FT dataset is different, and there is no obvious physical correlation between them in several cases. For example, PT with GV yields the best FT model with EBG, even though both properties are not directly correlated (Figure~{\ref{fig:heatmap-2}}b). Additionally, there is no symmetrical relationship between the datasets that constitute the best performing PT-FT pairs. For instance, PT with FE offers the best scores for GV800, but PT with GV does not yield the best scores for FE800. Ignoring FT on PZ, no PT dataset yields particularly worse performance than other datasets, indicating that when PT dataset size is capped, the specific property being trained on has little influence on FT. Among the FT datasets, we observe fairly good performance for GV, PH, and FE (R$^{2}$ score > 0.75), indicating that these properties are easier to generalize. We get average-performing models for BG, DC, and EBG (0.4 < R$^{2}$ score < 0.7), while PZ seems to be a particularly difficult dataset to FT or train from scratch (|R$^{2}$score| < 0.1).

\subsection{MPT model performance}
Utilizing the cumulative dataset of 132,270 points, we compare the performance of MPT models upon FT (using strategy 1 with additional MLPs) on the seven different properties versus scratch and the best pair-wise PT-FT models, as illustrated in Figure~{\ref{fig:heatmap-2}}c. The percentage change in performance (similar to the definition in Section~{\ref{sec:PT-size}}) of both MPT and pair-wise models versus scratch models are plotted as hashed and solid bars, respectively, in Figure~{\ref{fig:heatmap-2}}c. Blue (orange) bars indicate R$^2$ scores (MAEs), where positive (negative) values indicate an improved (worse) performance of MPT/pair-wise models compared to scratch. We have plotted percentage change in performance upon FT with PZ in Figure~S15b for ease of visualization. Note that during PT of MPT models, all points belonging to the 132,270 cumulative dataset is exposed, except the specific property (and the corresponding one-hot encoded vector) that is subsequently FT.

For the sake of comparison, we cap the FT (training) dataset size to 800 in Figure~{\ref{fig:heatmap-2}}c for both MPT and pair-wise (scratch) models, with Tables~S19 and S20 tabulating the results for other FT dataset sizes with MPT. Figure~S18f visualizes the results tabulated in Tables~S19 and S20. Using FT strategy 1, we maximize the size of the PT dataset that gives the best performance upon FT for each pair-wise model. For example, PT with FE and BG yield best performance upon FT with GV and DC, respectively (see Figure~{\ref{fig:heatmap-2}}b). Hence, we use FT(100K) and BG(50K) as PT dataset for FT with GV and DC and subsequently compare the obtained test R$^2$ and MAE scores with the MPT models. Similarly, we employ the full datasets of PH, EBG, FE, and GV during PT followed by FT on FE, BG, PZ, and EBG, respectively, to generate the best pair-wise models. In the case of PH, instead of choosing PZ for PT, we choose the next best model (i.e., BG(50K)) to increase the number of PT datapoints.

As displayed in Figures~{\ref{fig:heatmap-2}}c and S15b, the MPT model FT on a given property outperforms the corresponding scratch models in six out of seven cases. The lowest performance improvement with MPT models compared to scratch is observed in the case of the GV, with a $\sim$2.7\% increase. Additionally, the percentage of improvement in R$^{2}$ scores for the MPT model is higher than the corresponding best pair-wise PT-FT model in 3 out of 7 cases, while the improvement in MAEs for the MPT model is better than the pair-wise models in 5 out of 7 cases. The highest reduction in MAE (23.53\%), and the highest increase in R$^{2}$ scores (15.68\%) for MPT compared to best pair-wise models is observed for FT with PH and EBG, respectively.

Interestingly, the MPT model FT on FE shows negative transfer as the model shows lesser R$^{2}$ scores (and higher MAEs) with respect to the scratch model. We attribute this negative transfer behavior to the fact that the FE is the largest dataset in our work, which results in poor PT of the MPT model since a significant number of structures only have FE as the sole datapoint in our cumulative set ($\approx$26K), all of which are excluded during PT causing poor model generalization. Thus, the choice of large PT datasets, especially containing datapoints with at least one non-zero property, plays a significant role in training effective and generalizable MPT models. Excluding FT on FE, MPT models outperform best pair-wise models on R$^2$ scores in 3 out of 6 cases, and on MAEs in 5 out of 6 cases, highlighting their overall effectiveness when utilised for TL.

\subsection{MPT model on a completely unrelated dataset}
Given that accessing larger datasets by including multiple properties during PT improves model generalizability, we test our MPT framework on a task where the materials are out-of-distribution of those used used in the PT. We choose the JARVIS-DFT 2D dataset,\cite{choudhary2020joint} consisting of DFT-computed band gap for 1103 2D materials, as the out-of-distribution dataset. For this exercise, we PT a MPT model on all the seven properties combined, i.e., used all the datapoints of the cumulative 132,270 dataset. Further, we trained a scratch model on the 2D dataset and FT all best-size PT models (i.e., PT on FE(100K), BG(50K), PH, GV, and EBG), to compare the performance of the MPT model. We standardized and normalized the 2D dataset and split it into 90\% train and 10\% test sets, with the 90\% set used for FT (and training the scratch model) by employing FT strategy 1.

The test R$^2$ scores and MAEs for all models on the 2D dataset are tabulated in Table~{\ref{tab:unrelated}}. Importantly, the scores obtained from the MPT model is better than all pair-wise PT(best-size) and scratch models. The improvement in R$^{2}$scores and MAEs for the MPT model compared to scratch is 5.67\% and 15.54\%, respectively, and 6.27\% and 9.99\% on-average compared to the PT(best-size) models. The pair-wise model that exhibits the closest performance to the MPT model ($\sim$1.5\% deviation in MAE) is FE(100K), which is expected given that FE is the largest dataset among the seven considered within the MPT model. Thus, we observe that our MPT model can generalize well on datasets that are out-of-distribution to its PT datasets and indicates the employability of our MPT framework on other distinct material properties.  

\begin{table}[htbp]
\centering
\begin{tabular}{c@{\hspace{12mm}}c@{\hspace{10mm}}c@{\hspace{10mm}}c@{\hspace{10mm}}c@{\hspace{10mm}}}
\hline 
\hline \noalign{\vspace{1ex}}
\multicolumn{1}{c}{\textbf{PT models}} & \multicolumn{2}{c}{\textbf{R$^{2}$ score}} & \multicolumn{2}{c}{\textbf{MAE}}\\ 
\hline 
\hline \noalign{\vspace{1ex}}
\multicolumn{1}{c}{Scratch} & \multicolumn{2}{c}{ 0.635} & \multicolumn{2}{c}{0.148}
\\\hline \noalign{\vspace{1ex}}
\multicolumn{1}{c}{MPT} &\multicolumn{2}{c}{ 0.671} & \multicolumn{2}{c}{ 0.125}
\\\hline \noalign{\vspace{1ex}} 
\multicolumn{1}{c}{FE(100K)} &\multicolumn{2}{c}{ 0.670} & \multicolumn{2}{c}{ 0.127}
\\\hline \noalign{\vspace{1ex}}
\multicolumn{1}{c}{BG(50K)} &\multicolumn{2}{c}{ 0.617} & \multicolumn{2}{c}{ 0.138}
\\\hline \noalign{\vspace{1ex}}
\multicolumn{1}{c}{PH(1256)} &\multicolumn{2}{c}{ 0.628} & \multicolumn{2}{c}{ 0.145}
\\\hline \noalign{\vspace{1ex}}
\multicolumn{1}{c}{GV(10987)} &\multicolumn{2}{c}{ 0.626} & \multicolumn{2}{c}{ 0.143}
\\\hline \noalign{\vspace{1ex}}
\multicolumn{1}{c}{EBG(2481)} &\multicolumn{2}{c}{ 0.619} & \multicolumn{2}{c}{ 0.143}
\\\hline
\end{tabular}
\caption{Performance of scratch, MPT, and PT(best-size) models on the JARVIS-DFT 2D band gap dataset. MPT model considered here is trained on the full cumulative dataset comprising all seven properties considered. Both R$^2$ scores and MAEs listed are for the test dataset.}
\label{tab:unrelated}
\end{table}

\section{Discussion}
\label{sec:discussion}
In this work, we presented efficient methods for utilizing TL to deal with small dataset sizes that are typical of materials science. To optimize pair-wise TL in materials science, we chose seven different material property datasets spanning a size range of 941 to 132,752 datapoints. Using a GNN-based architecture, we considered different FT strategies, the influence of dataset sizes both in PT and FT, and optimized other hyperparameters. We found the PT-FT models to outperform scratch models, in terms of R$^2$ scores and MAEs, often while being FT on fewer datapoints. Importantly, we introduced a MPT model that was trained simultaneously on six out of the seven properties considered,wherein our MPT models performed better than both scratch and the best pair-wise models upon FT on several datasets. Also, the MPT model that was trained on all seven properties simultaneously offered the best performance upon FT on a completely out-of-distribution dataset consisting of DFT-computed 2D material band gaps. Our work provides foundational advancements in efficiently performing TL on several materials-based datasets.

On comparing our models with models from previous works, we observe a significant reduction in the number of datapoints required for PT and FT. For example, the MAEs obtained by Lee and Asahi\cite{lee2021transfer} for FE500 and BG500 are 0.149 and 0.866, respectively. Our models on the same dataset show a MAE of 0.057 and 0.142 for the same number of FT FE and BG datapoints, respectively, but utilising three orders of magnitude lower datapoints during PT (see Table~S21). Our FE(941)-PH800 shows an MAE that is 69.90\% lower than the MAE of the MOE model proposed by Chang et al.,{\cite{chang2022towards}} which used 18 properties for PT before FT on the PH dataset. Also, our MPT model PT on all properties but PH displays a decrease in MAE of 74.76\% compared to the MOE model. Thus, our combination of optimized FT strategies, hyperparameters, and the MPT framework outperform previous TL efforts, highlighting their applicability in extending to other materials datasets.

The distribution of datapoints in the seven datasets (Figure~{\ref{fig:alignn-data}}b) chosen for our study is different, with four datasets (BG, PZ, DC, and EBG) showing a highly skewed (i.e., log-normal) distribution than the others (GV, PH, and FE). Consequently, our models PT on PZ show poor performance even before FT, which may arise from the skewed nature of the dataset, possibly resulting in uneven representation within the train and test sets. Additionally, from our 7$\times$6 pair-wise TL models, we observe PT on normally distributed datasets yield the best FT performance for four out of seven cases (Figure~{\ref{fig:heatmap-2}}a). Also, incorporating larger amount of PT data with FE (i.e., from 50K to 100K datapoints) always resulted in better FT models, while going from 50K to 100K with a skewed BG dataset during PT did not always result in better FT models (Figure~{\ref{fig:ld-radar}}a and b). Hence, we expect TL models that are PT on normally distributed data to generally outperform models that are PT on skewed datasets. 

Apart from the hyperparameters optimized in this work (Section~{\ref{subsec:hyperparameter}}), batch size can be an important hyperparameter as well. We observed lower batch sizes to offer comparable or better performance for smaller dataset sizes ($<$1000) in pair-wise TL, motivating us to fix a batch size of 16 for smaller datasets. For larger datasets (>1000) in pair-wise TL, we used a batch size of 64, as reported in previous work.\cite{choudhary2021atomistic} We believe that we have used optimal batch sizes in our work since both pair-wise TL and MPT models outperform our scratch models. Nevertheless, further optimizations of batch size particularly for large datasets may improve performance. Note that model performance may also be improved by utilizing better FT strategies than those proposed in this work, such as combining strategies 1 and 2 (Figure~{\ref{fig:strategy}}).

In the case of the MPT framework, we observe significant variations in the model performance when different properties are combined during PT. A general trend is that both PT and FT losses tend to decrease with the increase in the number of properties that the model is trained on simultaneously, which is illustrated in Section~S4 and Figure~S14 of the SI. Therefore, it will be interesting to identify the best combinations of PT datasets, across a wider range than considered in this work, for constructing an optimal MPT model and improve knowledge transfer. 

All the predictions in our study are graph-level predictions, i.e., properties that depend on the (graphical representation of the) entire structure. Given that GNNs can also yield atomic (node) and bond (edge) level properties, it would be interesting to explore TL frameworks and strategies to predict properties at such levels (e.g., defect formation energies, site energies, bond dissociation energies, etc.). Another pathway to explore is the implementation of active learning to further improve the TL model performance on the target dataset by iteratively selecting and re-training on the most important instances of the dataset. The active learning strategy might be useful for target properties that are more scarce than those we have considered here. 

We expect our TL framework to be transferrable to other GNNs, including ones that exhibit more complex architectures than ALIGNN, such as NequIP\cite{batzner20223} and MACE.\cite{batatia2022mace} However, we expect GNNs that ignore critical structural information, such as bond-angles, to exhibit inferior performance than ALIGNN  while utilising similar TL strategies and hyperparameters. On the other hand, more complex GNN frameworks can be prone to over-fitting due to the higher number of parameters (and hyperparameters).\cite{omee2024structure} Thus, the TL framework proposed here may require modifications if more complex and deeper GNNs are utilised.

\section{Conclusion}
In conclusion, we provide an improved TL paradigm for effective knowledge transfer from source datasets to target datasets with a restricted amount of datapoints, which is highly relevant for materials science. By comparing generated R$^2$ scores and MAEs with scratch models, we rigorously investigated the impact of size of the FT and PT datasets and the FT strategy on the performance of pair-wise TL models. We observed pair-wise models to generally outperform scratch models across seven different materials property datasets. Additionally, we looked at training a model on several characteristics of the data at once and compared the performance of such MPT models versus both scratch and pair-wise models. In several cases, we found the MPT models to perform better (or similar) to the equivalent pair-wise models. Importantly, we observed the MPT model to perform significantly better than both scratch and pair-wise models upon FT on a 2D material dataset that was entirely out-of-distribution from the PT data, highlighting the effectiveness of our MPT framework.  With quantitative improvements in model performance, our GNN-based TL framework offers a comprehensive architecture that can lead to better predictions among data-scarce material property datasets at a low computational cost and accelerate materials discovery.  

\section*{Data and code availability}
All computed data, constructed models, and codes associated with this work are available online freely to all via our \href{https://github.com/sai-mat-group/transfer-learning-material-properties}{GitHub} repository.

\subsection*{Conflicts of Interest}
The Authors declare no competing financial or non-financial interests.

\section*{Acknowledgments}
G.S.G. and K.T.B. would like to acknowledge financial support from the Royal Society under grant number IES$\backslash$R3$\backslash$223036, and the United Kingdom Research and Innovation (UKRI) Engineering and Physical Sciences Research Council (EPSRC), under projects EP/Y000552/1 and EP/Y014405/1. G.S.G. acknowledges financial support from the Science and Engineering Research Board (SERB) of the Department of Science and Technology, Government of India, under sanction number IPA/2021/000007. R.D. thanks the Ministry of Human Resource Development, Government of India, for financial assistance. R.D. and G.S.G. acknowledge the computational resources provided by the Supercomputer Education and Research Centre, IISc, for enabling some of the calculations showcased in this work. We acknowledge National Supercomputing Mission (NSM) for providing computing resources of ‘Param Utkarsh’ at CDAC Knowledge Park, Bengaluru. PARAM Utkarsh is implemented by CDAC and supported by the Ministry of Electronics and Information Technology (MeitY) and Department of Science and Technology (DST), Government of India. Via our membership of the UK's HEC Materials Chemistry Consortium, which is funded by EPSRC (EP/X035859/1), this work used the ARCHER2 UK National Supercomputing Service (\href{http://www.archer2.ac.uk}{http://www.archer2.ac.uk}).



\bibliographystyle{unsrt}
\bibliography{sample}

\end{document}